\documentclass[a4paper,aps,preprint,prd,nofootinbib]{revtex4-1}
\usepackage{amsmath,amsthm}
\usepackage{comment,amssymb,latexsym,upref}
\usepackage{finalT}

\numberwithin{equation}{section}

\input lag.def

\begin{document}

\title{Lagrange coordinates for the Einstein-Euler equations}
\author{Todd A. Oliynyk}
\affiliation{School of Mathematical Sciences\\
Monash University, VIC 3800\\
Australia}
\email{todd.oliynyk@monash.edu}

\begin{abstract}
We derive a new symmetric hyperbolic formulation of the Einstein-Euler equations in Lagrange coordinates that
are adapted to the Frauendiener-Walton formulation of the Euler equations. As an application, we use this system to
show that the densitized lapse and zero shift coordinate systems for the vacuum Einstein equations are equivalent to
Lagrange coordinates for a fictitious fluid with a specific equation of state.
\end{abstract}

\maketitle 
\sect{intro}{introduction}

Perfect fluid balls are used to model many different types of physical objects such as gaseous planets and stars.
An important physical problem is to understand the evolution of
these fluid balls. In general relativity, this evolution
is governed by the Einstein-Euler equations\footnote{Lower case Greek indices (i.e. $\mu,\nu,\gamma$ )
will be run from $0$ to $3$ and will be used exclusively as coordinate indices.}
\lalign{EEa}{
R_{\mu\nu}  &= \kappa\bigl(T_{\mu\nu}-\Half T g_{\mu\nu}\bigr),   \label{EEa.1}\\
\nabla_\mu T^{\mu\nu}  &= 0, \label{EEa.2}
}
where $R_{\mu\nu}$ is the Ricci tensor of the Lorentzian metric
\eqn{met}{
g = g_{\mu\nu}dx^\mu dx^\nu
}
with signature $(-,+,+,+)$,
and
\eqn{stress}{
T^{\mu\nu} = (\rho+p)v^\mu v^\nu + p g^{\mu\nu}
}
is the stress energy tensor of a perfect fluid.
Here, $\rho$ is the proper energy density of the fluid, $p$ is the fluid pressure, and $v^\mu$ is the fluid four-velocity
normalized by\footnote{Following standard conventions, we will lower and raise the coordinate indices with
the metric $g_{\mu\nu}$  and inverse metrics $g^{\mu\nu}$, respectively.}
\eqn{vnorm}{
v_\mu v^\mu = -1.
}
Due to the presence of a free boundary at the
fluid vacuum interface, establishing existence and uniqueness of solutions to the equations \eqref{EEa.1}-\eqref{EEa.2}
is a difficult problem. This remains true even in the simpler settings where gravity is either Newtonian or absent.

A natural first step in studying the vacuum free boundary value problem is to use Lagrange coordinates so that the equations of motion can be formulated on
a fixed domain. Indeed, the first general local existence and uniqueness proof for solutions to the compressible non-relativistic Euler equations
representing a fluid ball used Lagrange coordinates \cite{Lind}.
However, the usual method of passing to Lagrange coordinates introduces technical complications due to the fact that the equations in the Lagrange representation are no longer of a standard hyperbolic form. Because of this, specialized techniques are needed to obtain existence. For example,
in \cite{Lind}, a Nash-Moser iteration technique was required. Other specialized techniques for handling the free boundary problem
for the compressible Euler equations can be found in \cite{CK1D,CK3D,JM,JM3D,Trak,Oli11}.

A second approach to the vacuum free boundary value problem initiated in \cite{Frie} (see also \cite{GRN})
is to derive a symmetric hyperbolic
form of the Einstein-Euler equations in Lagrange coordinates. In \cite{Frie,GRN}, the symmetric
hyperbolic system is derived by taking certain combinations of the the original variables along with their first
derivatives and writing them in Lagrange coordinates. Although this approach yields a symmetric hyperbolic system
formulated in Lagrange coordinates,
the boundary behavior of the system is much more difficult to analyze due to the addition of more variables,
which must also be controlled at the boundary. At this time,
it is not clear how or even if it is possible to use the symmetric hyperbolic systems of \cite{Frie} or \cite{GRN} to prove
the existence and uniqueness of solutions to the Einstein-Euler or even just Euler equations with a vacuum boundary.

In this article, our goal, as in \cite{Frie}, is to derive a symmetric hyperbolic system for the Einstein-Euler
equations in Lagrange coordinates. As in \cite{Frie}, we use a frame formulation for the Euler equations, but
instead of starting with the Euler equations in their standard presentation, we use a formulation due, independently,
to Frauendiener and Walton \cite{Frau,Walt}. In the Frauendiener-Walton formulation, the normalized four-velocity
$v^\mu$ and the proper energy density $\rho$ are combined into a single vector $w^\mu$.
Due to the vector nature of this formulation and its geometric structure,
it is possible to exploit the diffeomorphism freedom available to fix the Lagrange coordinates while retaining
a symmetric hyperbolic form for the Einstein-Euler equations. In particular, we are able to derive a symmetric
hyperbolic system without differentiating the Einstein-Euler equations. In principle, this should make the boundary
behaviour of the system we derive here easier to analyze than in \cite{Frie}. However, it remains to be seen
if our system is useful for studying the boundary value problem in the physical setting of 4 spacetime dimensions.
We are currently using this new system to investigate the boundary value problem in 4 spacetime dimensions, and while definitive results are not yet available,
we do note that the Lagrange coordinates introduced here have already proven useful in
2 spacetime dimensions to analyze the vacuum boundary problem \cite{Oli11}.

As an application of our Lagrange formulation, we show that it yields a symmetric hyperbolic formulation of the vacuum Einstein
equations in a densitized lapse and zero shift coordinate system. Distinct symmetric hyperbolic formulations of the vacuum Einstein
equations in a densitized lapse and zero shift coordinate systems have been derived by a number of different authors, for example
see \cite{Frie96,KST,NOR2004,SarTig}. What the results of this article show is that these coordinate systems are nothing more than different formulations of the vacuum Einstein equations in
Lagrange coordinates adapted to a fictitious fluid with a specific equation of state. Here we are using the term fictitious fluid to refer to
a relativistic fluid on spacetime that is not coupled to the Einstein equations via its stress energy tensor; the only purpose of
the fictitious fluid is to fix a coordinate system.

\sect{fw}{The Frauendiener-Walton formulation for the Euler equations}

In \cite{Frau,Walt}, Frauendiener and Walton independently showed that the isentropic Euler equations for a perfect fluid with an
equation of state of the form $p = p(\rho)$
can be written as
\leqn{eul1}{
A_{\mu\nu}{}^\gamma \nabla_\gamma w^\nu = 0,
}
where $w^\nu$ is a timelike vector field with norm
\eqn{eul2}{
w^2 = -w_\nu w^\nu > 0,
}
and
\leqn{eul3}{
A_{\mu\nu}{}^\gamma = \left(3 + \frac{1}{s^2}\right) \frac{w_\mu w_\nu}{w^2} w^\gamma + \delta^\gamma_\nu w_\mu
+\delta^\gamma_\mu w_\nu + w^\gamma g_{\mu\nu}.
}
We will refer to these equations as the \emph{Euler-Frauendiener-Walton (EFW)
equations}.

In the Frauendiener-Walton formulation, $s^2$ is a function of
\eqn{eul4}{
\zeta =\frac{1}{w},
}
where
\eqn{eul5}{
w= \sqrt{w^2}. }
An explicit formula for $s^2$ can be calculated in the following fashion (see \cite{Frau} for more details). First,
the pressure
$p=p(\zeta)$ is determined implicitly by the equation
\leqn{eul6}{
\zeta = \zeta_0\Phi(p(\zeta)),
}
where
\eqn{eul7}{
\Phi(p) = \exp\left(\int_{p_0}^p \frac{d\tilde{p}}{\rho(\tilde{p})+\tilde{p}}\right)
}
is the Lichnerowicz index of the fluid. From this, $s^2$ can be calculated using
the formula
\leqn{eul8}{
\frac{1}{s^2} = \left(\frac{\zeta f'(\zeta)}{f(\zeta)}-3 \right),
}
where
\eqn{eul9}{
f(\zeta) = \zeta^3 p'(\zeta).
}
Additionally, the proper energy density $\rho$ and the fluid velocity
$v^\mu$ can be recovered from
\leqn{eul10}{
\rho = \frac{f(\zeta)}{\zeta^2}-p(\zeta) \AND v^\mu = \zeta w^\mu.
}
As shown in \cite{Frau}, and also \cite{Walt}, the triple $\{\rho,p,v^\mu\}$ determined from \eqref{eul1}, \eqref{eul6}, and \eqref{eul10} satisfy the relativistic
Euler equations \eqref{EEa.2}.

\begin{rem} \label{fraurem}
In \cite{Frau} it is shown that the physical condition $s^2 <1$ (i.e. the speed of sound is less than the speed of light)
is sufficient to establish the positive definiteness of
the bilinear form
\leqn{equivB14}{
\ipe{f}{k} = - t_\gamma A_{\mu\nu}{}^\gamma \delta^{ij} f^{\mu}_i k^{\nu}_j
}
for any time like vector $t^\gamma$. This property along with the obvious symmetry $A_{\mu\nu}{}^\gamma=A_{\nu\mu}{}^\gamma$
(see \eqref{eul3}) is enough to guarantee that the system \eqref{eul1} is symmetric hyperbolic.
\end{rem}

\sect{harm}{The Einstein-Euler equations in harmonic coordinates}

The starting point for our derivation of a symmetric hyperbolic Lagrange formulation for the Einstein-Euler
equations is to assume that the coordinates $(x^\mu)$ are harmonic, that is
$g^{\mu\nu}\Gamma_{\mu\nu}^\gamma = 0$,
where it is known that the Einstein-Euler equations admit a symmetric hyperbolic formulation.
Our plan is to transfer this symmetric hyperbolic structure
to a Lagrange coordinate system.

\begin{rem} \label{harmrem}
In the following arguments, the harmonic gauge condition
can be replaced, without difficulty, by a generalized harmonic gauge condition
$g^{\mu\nu}(\Gamma_{\mu\nu}^\gamma -\hat{\Gamma}_{\mu\nu}^\gamma) = \hat{f}^\gamma$ where
$\hat{\Gamma}_{\mu\nu}^\gamma$ are the Christoffel symbols of a prescribed background metric
$\hat{g}_{\mu\nu}$, and $\hat{f}^\gamma$ is a prescribed vector field. In this article, we restrict ourselves
to standard harmonic coordinates for reasons of simplicity and the fact that there seems to be
little benefit for our purposes to consider more complicated gauges. However, in situations where
a boundary is present, it may become essential to use generalized harmonic gauges that are
adapted to the boundary. We plan to investigate these gauge issues in future work.
\end{rem}

In harmonic coordinates, the Einstein-Euler equations \eqref{EEa.1}-\eqref{EEa.2} are given by
\lalign{EEb}{
g^{\alpha\beta}\del{\alpha}\del{\beta} g_{\mu\nu} &= Q_{\mu\nu}\bigl(g_{\alpha\beta},\del{\gamma} g_{\alpha\beta}\bigr)-2\kappa\bigl(T_{\mu\nu}-\Half T g_{\mu\nu}\bigr),   \label{EEb.1}\\
\nabla_\mu T^{\mu\nu}  &= 0, \label{EEb.2}
}
where
\eqn{deldef}{
\del{\mu} = \frac{\partial\;}{\partial x^\mu},
}
and the map $Q(g,\del{} g)$ is quadratic in $\del{} g$ and analytic in $(g,\del{} g)$ with $g$ non-degenerate. To write these equations
as a symmetric hyperbolic system, we use a first order formulation for the gravitational field by defining (see \cite{Geroch} or \cite[\S II]{KRSW})
\leqn{gvarsdef}{
g_{\gamma\mu\nu} := \del{\gamma} g_{\mu\nu} \AND  B^{\alpha \beta \gamma} := -u^\alpha g^{\beta \gamma} - u^\beta g^{\alpha\gamma}
+ u^\gamma g^{\alpha\beta},
}
where $u^\nu$ is any time like vector field. For the Euler equations, we use the Frauendiener-Walton formulation described in Section \ref{fw}.
With the definitions \eqref{gvarsdef} and the Frauendiener-Walton formulation, we can, after letting
\eqn{uwdef}{
u^\mu=w^\mu,
} write the Einstein-Euler system \eqref{EEb.1}-\eqref{EEb.2}
as
\lalign{EEc}{
B^{\alpha \beta \gamma}\del{\gamma} g_{\beta\mu\nu} &= -w^\alpha\bigl(Q_{\mu\nu}(g_{\tau\delta},g_{\sigma\tau\delta})
-2\kappa\bigl(T_{\mu\nu}-\Half T g_{\mu\nu}\bigr)\bigr) \label{EEc.1}, \\
A_{\mu\nu}{}^\gamma \nabla_\gamma w^\mu & = 0 \label{EEc.2}, \\
w^\gamma \del{\gamma} g_{\mu\nu} &= w^\gamma g_{\gamma\mu\nu} \label{EEc.3}
}
where
\eqn{Tfw}{
T^{\mu\nu} = \left(\rho\left(\frac{1}{w}\right)+p\left(\frac{1}{w}\right)\right)\frac{w^\mu w^\nu}{w^2} + p\left(\frac{1}{w}\right)g^{\mu\nu}.
}
Here, $\rho$ and $p$ are determined by the formulas \eqref{eul6} and \eqref{eul10}, respectively.
We also note that in this formulation the Chritoffel symbols are calculated using
\eqn{Christ}{
\Gamma^\gamma_{\mu\nu} = \Half g^{\gamma\sigma}\bigl(g_{\mu\nu\sigma} + g_{\nu\mu\sigma}- g_{\sigma\mu\nu}\bigr).
}

The system \eqref{EEc.1}-\eqref{EEc.3} is symmetric hyperbolic, and so, given suitable initial data, local existence and
uniqueness follows by standard theory.  Suitable initial data is chosen by prescribing the fields
\eqn{idata}{
\bigl(g_{\mu\nu}|_{\Sigma},g_{\gamma\mu\nu}|_{\Sigma}=\del{\gamma}g_{\mu\nu}|_{\Sigma},w^\mu|_{\Sigma}\bigr)
}
on a spacelike hypersurface $\Sigma$ with unit conormal $n_\mu$ where
\leqn{widata}{
\text{$w^\mu|_{\Sigma}$ is timelike,}
}
and the constraint equations (see \cite[\S 10.2]{Wald} or  \cite[Ch. VII]{CB})
\lalign{ceqns}{
\bigl(G_{\mu\nu}n^\nu -\kappa T_{\mu\nu}n^{\nu}\bigr)|_{\Sigma} &=0 &&\text{(Gravitational constraint equations),} \label{ceqns.1}
\intertext{and}
g^{\mu\nu}\Gamma_{\mu\nu}^\gamma|_{\Sigma} &= 0
&& \text{(harmonic constraints)} \label{ceqns.2}
}
are satisfied. Also, we must choose the initial
data $w^\mu|_{\Sigma}$ so that energy density and sound speed determined by $w^\mu$ (see \eqref{eul8} and \eqref{eul10}) satisfies
\leqn{irho}{
\rho|_{\Sigma} > 0 \AND s^2|_{\Sigma} < 1.
}

Finally, to simplify the discussion below, we assume that our harmonic coordinates are chosen so that
\leqn{Sigma}{
\Sigma = \{\, (0,x^I) \,|\, (x^I)\in \Sigmah
 \}.
}
Since we will not be addressing the boundary value problem, we will not, by the finite
propagation speed property of hyperbolic equations, lose anything in assuming that
\eqn{Sigmah}{
\Sigmah = \Tbb^3,
}
and the $(x^I)$ are the standard period coordinates on $\Tbb^3$ with period $1$. 
\sect{lagcoords}{Lagrange coordinates}

Traditionally, Lagrange coordinates are introduced by using the flow map of
the normalized four-velocity $v=v^\mu\del{\mu}$ to define coordinates that
trivialize $v$, that is the components of $v$ are given by
$v^\mu = \delta^\mu_0$. However, in the Frauendiener-Walton formulation,
it is more natural to introduce coordinates that trivialize the vector field $w=w^\mu\del{\mu}$. For obvious
reasons, we will also refer to these coordinates as Lagrange coordinates.

\subsect{triv}{Trivializing $w=w^\mu\del{\mu}$}

The first step in constructing the Lagrange coordinates is to let
\leqn{flowA}{
\Fc_\tau(x^\nu) = \bigl(\Fc^\mu_{\tau}(x^\nu)\bigr)
}
denote the flow map of the vector field $w=w^\mu\del{\mu}$, that is $\Fc^\mu_{\tau}(x^\nu)$ is the unique
solution to the initial value problem
\alin{flowB}{
\frac{d\;}{d\tau} \Fc^\mu_\tau(x^\nu) & = w^\mu(x^\nu),\\
\Fc^\mu_0(x^\nu) &= x^\mu.
}
We use this flow map to define a new set of coordinates $(\xb^\mu)$ via the formula\footnote{Upper case Greek indices
(i.e. $\Lambda, \Omega, \Gamma$) will always run from $1$ to $3$.}
\leqn{phidefA}{
(x^\mu) = \phi(\xb^\mu) := \Fc_{\xb^0}(0,\xb^\Lambda).
}
Since this diffeomorphism is generated by the flow of $w$, the pullback of $w$ satisfies
\leqn{wbdefA}{
\wb := (\phi^* w) = \delb{0}
}
where
\eqn{delbdef}{
\delb{\mu} = \frac{\partial\;}{\partial \xb^\mu}.
}
This shows that the coordinates $(\xb^\mu)$ do in fact define a Lagrange coordinate system.

Letting
\eqn{Jdef}{
J^\mu_\nu := \delb{\nu}\phi^\mu
}
denote the Jacobian matrix of the coordinate transformation \eqref{phidefA}, and
\leqn{Jinv}{
(\Jch^\mu_\nu) := (J^\mu_\nu)^{-1}
}
its inverse, the formula
\eqn{wpback}{
\wb^\mu = \Jch^\mu_\nu w^\nu\circ\phi
}
for the pullback shows that \eqref{wbdefA}
can also be written as
\leqn{wbdefB}{
\delb{0}\phi^\mu = w^\mu \circ\phi.
}

\subsect{Jacobian}{Evolution of the inverse Jacobian matrix}

\subsubsect{frameEul}{A frame formulation for the Euler equations}

The next step in transforming the equations \eqref{EEc.1}-\eqref{EEc.3} into the Lagrange coordinates is to derive appropriate
evolution equations so that we can control the inverse Jacobian matrix \eqref{Jinv}. We begin this task by introducing
a frame\footnote{We reserve lower case Latin indices (i.e. $i,j,k$) for frame indices that run from $0$ to $3$. The frame
and coordinate indices will only coincide when the frame happens to be defined by a coordinate basis.}
\eqn{eidef}{
e_i = e^\mu_i\del{\mu} \qquad (i=0,1,2,3)
}
where
\leqn{e0def}{
e_0 := w,
}
and the remaining vectors\footnote{Upper case Latin indices
(i.e. $I, J, K$) will always run from $1$ to $3$.} $\{e_I\}_{I=1}^3$ are determined by Lie transport equations
\leqn{eIdefa}{
[e_0,e_I] = 0.
}
Our motivation for this choice of evolution for the frame vectors $\{e_I\}_{I=1}^3$ is that Lie transport
behaves naturally under the action
of diffeomorphisms, and in particular the transformation to Lagrange coordinates.

Although the combined evolution system \eqref{eul1} and \eqref{eIdefa} for
the frame $e_i^\mu$ coefficients is not symmetric hyperbolic, it is still possible to prove existence
of solutions. To see this, we recall that \eqref{eul1} is symmetric hyperbolic, and therefore,
given appropriate initial data $e^\mu_0|_{\Sigma}$ and assuming $g_{\mu\nu}$ has the required differentiability, standard local existence and
uniqueness theorems guarantee the existence of a solution $w^\mu = e^\mu_0$ to \eqref{eul1}. We can then solve \eqref{eIdefa}, written explicitly as
\eqn{eIdefrem}{
e^\mu_0 \del{\mu} e_I^\nu - (\del{\nu} e_0^\mu)e_I^\nu = 0,
}
for given initial data $e^\mu_I|_{\Sigma}$ either by the method of characteristics\footnote{Or in other words by propagating the
initial data $e_I|_{\Sigma}$ by the flow of $e_0$.} or by treating it as a symmetric hyperbolic system
with coefficients $e^\mu_0$ and $\del{\nu} e_0^\mu$ determined by the solution $e^\mu_0$. Of course, using either of these methods
it seems as though there is a loss of differentiability in the sense that even if the initial data
$e^\mu_0|_{\Sigma}$ and $e^\mu_I|_{\Sigma}$ have the same regularity, the $e^\mu_I$ generated from this initial data appear to
have one less order of regularity compared to the $e^\mu_0$. This apparent loss of differentiability arises because the
system \eqref{eul1} and \eqref{eIdefa} is not symmetric hyperbolic with respect to the frame coefficients.
Remarkably, as we shall see below, the structure of
the Euler-Frauendiner-Walton equations \eqref{eul1} guarantees that \eqref{eIdefa} is equivalent to
a symmetric hyperbolic equation for the $e_I^\mu$ whose coefficients do not involve derivatives of the frame components $e_j^\mu$ that,
in turn, implies that no loss of differentiability actually occurs.

Letting
\leqn{theta1}{
\theta^i = \theta^i_\mu dx^\mu \qquad (\theta^i_\mu e^\mu_j = \delta^i_j)
}
denote the coframe, we recall that the connection coefficients $\omega_i{}^k{}_j$ are defined
by
\eqn{omegadef1}{
\nabla_{e_i} e_j = \omega_i{}^k{}_j e_k
}
where $\nabla$ is the Levi-Civita connection of $g_{ij}$.
We define the connection 1-forms $\omega^k{}_j$ in the standard fashion
\eqn{omegadef2}{
\omega^k{}_j = \omega_i{}^k{}_j\theta^i.
}
We also set\footnote{In this article, we will follow standard convention and lower and raise the frame indices (i.e. $i,j,k$) with the
frame and inverse frame metrics $g_{ij}$ and $g^{ij}$, respectively.}
\eqn{omegadef3}{
\omega_{kj} = g_{kl}\omega^l{}_j = \omega_{ikj}\theta^i
}
where
\leqn{fmet}{
g_{ij} := g(e_i,e_j) = g_{\mu\nu}e_i^\mu e_j^\nu,
}
is the frame metric
and
\leqn{omegadef4}{
\omega_{ikj} = g_{kl}\omega_i{}^l{}_j = g(\nabla_{e_i} e_j, e_k).
}

For the evolution equation \eqref{eIdefa} to produce vector fields $\{e_I\}_{I=1}^3$ that are useful for our purposes, we need to partially restrict
the choice of initial data $\{e_I^\mu|_{\Sigma}\}_{I=1}^3$ beyond requiring that the $\{e_i|_{\Sigma}\}_{i=0}^3$ are linearly independent.
In order to describe this restriction, we introduce a function $F(\zeta)$ defined by
\leqn{FdefA}{
F(\zeta) = F_0 \exp\left(-\int^\zeta_{\zeta_0} \frac{1}{\eta s^2(\eta)}\, d\eta\right)
}
where $F_0,\zeta_0$ are arbitrary positive constants.
By design, $F(\zeta)$ satisfies the
differential equation
\leqn{FdefB}{
F'(\zeta) = -\frac{F(\zeta)}{\zeta s^2(\zeta)}.
}
We also note that
\leqn{w2}{
w^2 = -g_{00} \AND \zeta = \frac{1}{w} = \left(\frac{1}{-g_{00}}\right)^{\frac{1}{2}}
}
since $e_0=w$.

We are now ready to show that the equations \eqref{eul1} and \eqref{eIdefa} used to evolve the frame $\{e_i\}_{i=0}^3$
are equivalent to a symmetric hyperbolic system. It is worth noting that the following Proposition does not rely on
the metric $g_{\mu\nu}$ satisfying the Einstein equations, rather it is a statement about the
EFW equations \eqref{eul1} that is valid for arbitrary metrics.
\begin{prop} \label{framesymA}
Suppose $T>0$, $U_T = (0,T)\times \Sigmah$, $g_{\mu\nu}\in C^1(U_T)$, and $e_0=w$ and
$\{e_I\}_{I=1}^3$ are $C^1$ solutions of
\leqn{eul1a}{
A_{\mu\nu}{}^\gamma \nabla_\gamma w^\nu = 0,
}
\leqn{eIdef}{
[e_0,e_I] = 0,
}
on $U_T$, respectively. If the $\{e_i|_{\Sigma}\}_{i=0}^3$ are linearly independent and
satisfy
\leqn{eIidata}{
g(e_0,e_I)|_{\Sigma} = 0 \AND \det\bigl(g(e_I,e_J)|_{\Sigma}\bigr)= F\bigl((-g(e_0,e_0)|_{\Sigma})^{-\frac{1}{2}}\bigr)^2,
}
then the frame $\{e_i\}_{i=0}^3$ satisfies
\leqn{fwev11}{
g(e_0,e_J)= 0 \AND F\bigl( (-g(e_0,e_0))^{-1/2}\bigr)^2 = \det(g(e_I,e_J))
}
on $U_T$, and defines a $C^1$ solution of the symmetric hyperbolic system
\lalign{fwev21}{
A_{\nu\mu}{}^\gamma \nabla_\gamma e^\mu_l &= -2 g(e_0,e_0) \theta^i_\nu \delta^0_{(i}\delta^k_{j)}\sigma_l{}^j{}_k, \label{fwev21.1}\\
e_0(\sigma_l{}^j{}_k) & = 0 \label{fwev21.2}
}
on $U_T$ where
\leqn{fwev22}{
\sigma_l{}^j{}_k = \theta^j([e_l,e_k]) = \theta^j_\lambda
\bigl( e^\sigma_l \del{\sigma} e^\lambda_k  - e^\sigma_k \del{\sigma} e^\lambda_l \bigr)
}
and
\leqn{fwev22a}{
\sigma_0{}^j{}_k = \sigma_k{}^j{}_0 = 0.
}
\end{prop}
\begin{proof}
Letting
\leqn{Adef}{
A^{ijk} = g^{ip}g^{jq}e^\mu_p e^\nu_q \theta^k_\gamma A_{\mu \nu}{}^\gamma,
}
a short calculation using \eqref{eul3}, \eqref{e0def}, and \eqref{fmet} shows
that
\leqn{Adef2}{
A^{ijk}
= \left(3+\frac{1}{s^2}\right)\frac{\delta^i_0\delta^j_0\delta^k_0}{-g_{00}} + \delta^i_0 g^{jk} + g^{ik}\delta^j_0
+g^{ij}\delta^k_0 .
}
Moreover, it follows from \eqref{e0def} and \eqref{omegadef4} that
\leqn{omegadef5}{
\omega_{kj0} = e_k^\mu \nabla_\mu w^\nu e_j^\gamma g_{\nu\gamma}.
}
Together, \eqref{Adef2}, \eqref{omegadef5} and the invertibility of $g^{ij}$ and $e^\mu_i$ show that the EFW equations \eqref{eul1a} are equivalent to
\leqn{fwev1}{
A^{ijk}\omega_{kj0} = 0.
}

Since the connection $\omega^{k}{}_j$ is torsion free\footnote{This follows by virtue of $\omega^i{}_j$ being the Levi-Civita
connection of $g_{ij}$.}, it satisfies the  Cartan structure equation (see \cite[Ch.V,\S B]{Choq})
\eqn{cframe1}{
\ed \theta^i + \omega^i{}_j\Wp \theta^j = 0,
}
or equivalently
\leqn{cframe2}{
[e_i,e_j] = \bigl(\omega_i{}^k{}_j - \omega_j{}^k{}_i\bigr)e_k.
}
This allows us to write the evolution equation \eqref{eIdef} as
\eqn{eIev}{
\omega_{0jI} = \omega_{Ij0}.
}
Using this and \eqref{Adef2}, a short calculation shows that \eqref{fwev1} is equivalent to the following equations
\lalign{fwev2}{
\left(\left(3+\frac{1}{s^2}\right)\frac{1}{g_{00}}-3g^{00}\right)\omega_{000}
-g^{IJ}\omega_{0IJ} - 2g^{0J}\bigl(\omega_{00J}+\omega_{0J0}\bigr) &=0,\label{fwev2.1}\\
2g^{I0}\omega_{000} + g^{IJ}\bigl(\omega_{00J}+ \omega_{0J0}\bigr) & = 0. \label{fwev2.2}
}

The Cartan structure equation
\eqn{cframe3}{
\ed g_{ij} = \omega_{ij}+\omega_{ji},
}
or equivalently
\leqn{cframe4}{
e_{k}(g_{ij}) = \omega_{kij}+\omega_{kji},
}
implies that
\eqn{cframe5}{
e_0(g_{0J}) = \omega_{00J}+\omega_{0J0},
}
which allows us to write \eqref{fwev2.2} as
\leqn{fwev3}{
g^{IJ}e_0(g_{0J}) + 2 g^{I0}\omega_{000} = 0.
}

Writing the frame metric in matrix form
\eqn{gmat}{
(g_{ij}) = \begin{pmatrix}g_{00}& g_{0J}\\ g_{I0} & g_{IJ}\end{pmatrix},
}
the inverse frame metric is given by
\eqn{invgmet}{
(g^{ij}) = \begin{pmatrix}\begin{displaystyle}\frac{1}{g_{00}-g_{L0}\gch^{LM}g_{0M}} \end{displaystyle} &
\begin{displaystyle} -\frac{g_{0M}\gch^{MJ}}{g_{00}-g_{L0}\gch^{LM}g_{0M}}\end{displaystyle}\\
\begin{displaystyle}-\frac{g_{0L}\gch^{IL}}{g_{00}-g_{L0}\gch^{LM}g_{0M}} \end{displaystyle} &
\begin{displaystyle} \gch^{IJ}+\frac{\gch^{IL}g_{L0} g_{M0}\gch^{MJ}}{g_{00}-g_{L0}\gch^{LM}g_{0M}}\end{displaystyle} \end{pmatrix}
}
where $\gch^{IJ}$ is the matrix inverse of $g_{IJ}$. Using this formula, we can write \eqref{fwev3} as
\eqn{fwev4}{
e_0(g_{0J}) - \left(\frac{2\omega_{000} \gch_{JI}\gch^{IL}}{g_{00}-g_{L0}\gch^{LM}g_{0M}}\right)g_{0L} = 0
}
where $\gch_{IJ}$ is the matrix inverse of $g^{IJ}$. Viewing this as an evolution equation for $g_{0J}$, it
follows directly from \eqref{eIidata} that
\leqn{fwev5}{
g_{0J} = 0.
}
Substituting this into \eqref{fwev2.1} then yields
\leqn{fwev6}{
\frac{1}{s^2 g_{00}}\omega_{000} - g^{IJ}\omega_{0IJ} = 0
}
where in deriving this we have used the fact that \eqref{fwev5} implies that
\leqn{fwev5a}{
(g^{IJ})=(g_{IJ})^{-1}, \quad g^{00}=\frac{1}{g_{00}} \AND g^{0J} = 0.
}
We also see that
\leqn{fwev7}{
\omega_{l0J}+ \omega_{lJ0} = 0
}
follows directly from \eqref{fwev5} and \eqref{cframe4}.

Next, we observe that
\lalign{fwev8}{
\Half e_{l}\bigl[\ln\bigl(\det(g_{IJ})\bigr)\bigr] & = \Half g^{IJ} e_{l}(g_{IJ}) &&
\text{(by \eqref{fwev5a})} \notag\\
& = g^{IJ}\omega_{lIJ} &&\text{(by \eqref{cframe4})}, \label{fwev8.1}
}
and
\lalign{fwev9}{
\Half e_l\bigl[\ln\bigl[ F\bigl( (-g_{00})^{-1/2}\bigr)^2\bigr]\bigr] & =  -\frac{1}{2}
\begin{displaystyle}\frac{F'\bigl( (-g_{00})^{-1/2}\bigr)}{F\bigl( (-g_{00})^{-1/2}\bigr)} \end{displaystyle}
\begin{displaystyle}\frac{e_{l}(g_{00})}{(-g_{00})^{3/2}} \end{displaystyle}\notag \\
& = \frac{e_l(g_{00})}{2 s^2 g_{00}}&& \text{(by \eqref{FdefB} and \eqref{w2})} \notag\\
& = \frac{1}{s^2 g_{00}} \omega_{l00} && \text{(by \eqref{cframe4})}. \label{fwev9.1}
}
Setting $l=0$ in \eqref{fwev8.1} and \eqref{fwev9.1}, it follows from \eqref{fwev6} that
\eqn{fwev10}{
e_0\left(\ln\left(\frac{F\bigl( (-g_{00})^{-1/2}\bigr)^2}{\det(g_{IJ})}\right)\right) = 0,
}
and hence that
\eqn{fwev11a}{
F\bigl( (-g_{00})^{-1/2}\bigr)^2 = \det(g_{IJ})
}
by \eqref{eIidata}. Taking logarithm of the square of both sides of this expression and then differentiating yields
\eqn{fwev12}{
\Half e_l\bigl[\ln\bigl[ F\bigl( (-g_{00})^{-1/2}\bigr)^2\bigr]\bigr] = \Half e_{l}\bigl[\ln\bigl(\det(g_{IJ})\bigr)\bigr],
}
and this implies, with the help of \eqref{fwev8.1} and \eqref{fwev9.1}, that
\leqn{fwev13}{
\frac{1}{s^2 g_{00}} \omega_{l00} - g^{IJ} \omega_{lIJ} = 0.
}

Next, we observe, using \eqref{Adef2}, \eqref{fwev5}, and \eqref{fwev5a}, that the equations
\eqref{fwev7} and \eqref{fwev13} are equivalent to
\leqn{fwev14}{
A^{ijk}\omega_{ljk} = 0.
}
By the Jacobi identity
\eqn{Jac}{
[e_0,[e_I,e_J]]+[e_J,[e_0,e_I]]+[e_I,[e_J,e_0]] =0
}
and the evolution equation \eqref{eIdef}, we have that
\eqn{fwev15}{
[e_0,[e_I,e_J]]= 0,
}
and this implies that
\leqn{fwev16}{
e_0(\omega_I{}^k{}_J-\omega_J{}^k{}_I) = 0
}
by \eqref{eIdef} and \eqref{cframe2}. We also note that
\leqn{fwev17}{
\omega_0{}^j{}_I-\omega_I{}^j{}_0 = 0
}
follows from \eqref{eIdef} and \eqref{cframe2}.  Defining
\leqn{sigdef}{
\sigma_i{}^k{}_j = \omega_i{}^k{}_j-\omega_j{}^k{}_i,
}
we get from \eqref{fwev16} and \eqref{fwev17} that
\leqn{fwev18}{
e_0(\sigma_i{}^j{}_k)=0 \AND \sigma_0{}^j{}_k=\sigma_k{}^j{}_0 = 0.
}

Writing \eqref{fwev14} as
\eqn{fwev19}{
A^{ijk}\omega_{kjl} + A^i{}_j{}^k\bigl(\omega_l{}^j{}_k-\omega_k{}^j{}_l\bigr)=0,
}
we see from \eqref{fwev18} that
\leqn{fwev20}{
A^{i}{}_{j}{}^k\omega_{l}{}^j{}_k = \bigl(\delta^i_0\delta^k_j+g^{ik}g_{j0}\bigr)\sigma_l{}^j{}_k.
}
Using (see \eqref{omegadef4})
\eqn{cframe6}{
\omega_{kjl} = g_{\mu\tau} e^\mu_je^\gamma_k\nabla_\gamma e^\tau_l,
}
\eqref{theta1}, \eqref{Adef}, and  \eqref{fwev5}, we can transform \eqref{fwev20} to a coordinate basis to get
\eqn{fwev21a}{
A_{\nu\mu}{}^\gamma \nabla_\gamma e^\mu_l = 2 w^2 \theta^i_\nu \delta^0_{(i}\delta^k_{j)}\sigma_l{}^j{}_k
}
where we note that
\eqn{fwev22b}{
\sigma_l{}^j{}_k = \bigl( \theta^j([e_l,e_k]) \bigr) = \Bigl(\theta^j_\lambda
\bigl( e^\sigma_l \del{\sigma} e^\lambda_k  - e^\sigma_k \del{\sigma} e^\lambda_l \bigr)\Bigl).
}
This completes the proof.
\end{proof}

We now turn to showing that the system \eqref{fwev21.1}-\eqref{fwev21.2} is equivalent to \eqref{eul1a}-\eqref{eIdef}.

\begin{prop} \label{equivA}
Suppose $T>0$, $U_T = (0,T)\times \Sigmah$, $g_{\mu\nu}\in C^1(U_T)$, and
$\{\{e_i\}_{i=0}^3,\sigma_{l}{}^j{}_k\}$ is a $C^2$ solution of \eqref{fwev21.1}-\eqref{fwev21.2} that
satisfies the initial data constraints
\lgath{equivA1}{
[e_0,e_I]|_{\Sigma} = 0, \quad g_{0I}|_{\Sigma} = 0, \quad \det\bigl(g_{IJ}|_{\Sigma}\bigr)= F\bigl((-g_{00}|_{\Sigma})^{-\frac{1}{2}}\bigr)
\label{equivA1.1}
\intertext{and}
\sigma_l{}^j{}_k|_{\Sigma} = \bigl( \theta^j([e_l,e_k]) \bigr)|_{\Sigma} = \Bigl(\theta^j_\lambda
\bigl( e^\sigma_l \del{\sigma} e^\lambda_k  - e^\sigma_k \del{\sigma} e^\lambda_l \bigr)\Bigl)\bigl|_{\Sigma}. \label{equivA1.2}
}
Then $\{e_i\}_{i=0}^3$ with $w=e_0$ defines a $C^2$ solution of \eqref{eul1a}-\eqref{eIdef} on $U_T$.
\end{prop}
\begin{proof}
Let $w=e_0$, and suppose that $\{\{e_i\}_{i=0}^3,\sigma_{l}{}^j{}_k\}$ is a $C^2$ solution of \eqref{fwev21.1}-\eqref{fwev21.2} and that
the constraints on the initial data \eqref{equivA1.1}-\eqref{equivA1.2} are satisfied. Then
\leqn{equivA2}{
\sigma_0{}^j{}_k = 0
}
by the the evolution equation \eqref{fwev21.2} and the fact that the conditions on the
initial data \eqref{equivA1.1}-\eqref{equivA1.2} imply that $\sigma_0{}^j{}_k|_{\Sigma}=0$.
Substituting \eqref{equivA2} into the evolution equation \eqref{fwev21.1} then yields
\leqn{equivA3}{
A_{\nu\mu}{}^\gamma \nabla_\gamma e_0^\mu = 0,
}
or in other words, the vector field $w = e_0$ satisfies the EFW equations \eqref{eul1}.
Setting $\ech_0= e_0$, we can propagate the initial data $e_I|_{\Sigma}$ by the flow of $\ech_0$ to
get a $C^1$ solution $\{\ech_j\}_{j=0}^3$ on $U_T$ of the initial value problem
\lalign{equivA4}{
\Achk_{\nu\mu}{}^\gamma \nabla_\gamma \ech_0 & = 0, \label{equivA4.1} \\
[\ech_0,\ech_I] & = 0, \label{equivA4.2}\\
\ech_j|_{\Sigma} & = e_j|_{\Sigma} \label{equivA4.3}
}
where
\leqn{equivA5}{
\Achk_{\nu\mu}{}^\gamma = \left(3 + \frac{1}{\sch^2}\right) \frac{\wch_\mu \wch_\nu}{\wch^2} \wch^\gamma + \delta^\gamma_\nu \wch_\mu
+\delta^\gamma_\mu \wch_\nu + \wch^\gamma g_{\mu\nu},
}
$\wch = \ech_0$, $\sch^2 = s^2((\wch^2)^{-1/2})$ and $\wch^2 = \wch_\mu \wch^\mu$.

Since $\ech_0 = w$, it follows from \eqref{equivA1.1}, \eqref{equivA4.1}-\eqref{equivA5} and Proposition \ref{framesymA}
that $\{\{\ech_i\}_{i=0}^3,\sigmach_l{}^j{}_k\}$, where $\sigmach_l{}^j{}_k = \thetach^j([\ech_l,\ech_k])$,
defines a $C^1$ solution of \eqref{fwev21.1}-\eqref{fwev21.2}. Therefore $\{\{\ech_i\}_{i=0}^3,\sigmach_l{}^j{}_k\}$
and $\{\{e_i\}_{i=0}^3,\sigma_l{}^j{}_k\}$ both define $C^1$ solutions of \eqref{fwev21.1}-\eqref{fwev21.2} on $U_T$ with the same initial data.
By the uniqueness of $C^1$ solutions of symmetric hyperbolic differential equations, we conclude
that $\ech_i = e_i$ on $U_T$ for $i=0,1,2,3$, and the proof is complete.
\end{proof}

\subsubsect{bequiv}{Equivalence of \eqref{eul1a}-\eqref{eIdef} and \eqref{fwev21.1}-\eqref{fwev21.2} in the presence of a
boundary}

Because we are not directly confronting the initial boundary value problem in this article, the results of this section are
not used in the remainder of this article. However, they do represent a step towards the analysis
of the initial boundary value problem. What we accomplish in this section is to establish the equivalence of the
two hyperbolic systems \eqref{eul1a}-\eqref{eIdef} and \eqref{fwev21.1}-\eqref{fwev21.2} in the presence of
a boundary where the free boundary condition
\leqn{fbcA}{
p|_{\Gamma} = 0
}
is satisfied. Here, $\Gamma$ is the vacuum boundary, which separates the regions of positive fluid energy density $\rho > 0$
from the vacuum where $\rho = 0$.

Before proceeding, we recall that $p = p(\zeta)$ where $\zeta = 1/w$, and this implies
that the boundary condition \eqref{fbcA} is equivalent to
\leqn{fbcB}{
g(e_0,e_0)|_{\Gamma} = c_0
}
where $c_0 \in \Rbb_{<0}$ satisfies
\eqn{fbcC}{
p\bigl((-c_0)^{-1/2}\bigr) = 0
}
and as above, $e_0=w$. Next, we let $\Sigmah_0 \subset \Sigmah$ be a open subset diffeomorphic to a ball and
we let $\partial \Sigmah_0$ denote its boundary. For any $\tau \in \Rbb$, we define
\leqn{Vdef}{
V_\tau =  \bigcup_{0\leq t \leq \tau} \Sigma_t
}
where $\Sigma_t = \Fc_t\bigl(\{0\}\times \Sigmah_0\bigr)$ and $\Fc_t$ is, as above, the
flow of $e_0=w$. Spacetime regions of the form \eqref{Vdef} are precisely  the kind
where one solves the free boundary vacuum problem with boundary condition \eqref{fbcA}, or
equivalently, \eqref{fbcB} on the boundary region
\eqn{Gammadef}{
\Gamma_\tau = \bigcup_{0\leq t \leq \tau}\Fc_t\bigl(\{0\}\times \partial \Sigmah_0\bigr).
}

From the inspection of the proof of  Proposition \eqref{framesymA}, it is clear that boundary
conditions do not play a role in the proof and so the following result also holds.
\begin{prop} \label{framesymB}
Suppose $T>0$, $g_{\mu\nu}\in C^1(V_T)$, and $e_0=w$ and
$\{e_I\}_{I=1}^3$ are $C^1$ solutions of the boundary value problem
\lalign{framesymB1}{
A_{\mu\nu}{}^\gamma \nabla_\gamma w^\nu &= 0, \label{framesymB1.1} \\
[e_0,e_I] & = 0,  \label{framesymB1.2}\\
g(e_0,e_0)|_{\Gamma_T} & = c_0, \label{framesymB1.3}
}
on $V_T$, respectively. If the $\{e_i|_{\Sigma}\}_{i=0}^3$ are linearly independent and
satisfy
\eqn{framesymB2}{
g(e_0,e_I)|_{\Sigma_0} = 0 \AND \det\bigl(g(e_I,e_J)|_{\Sigma_0}\bigr)= F\bigl((-g(e_0,e_0)|_{\Sigma_0})^{-\frac{1}{2}}\bigr)^2,
}
then the frame $\{e_i\}_{i=0}^3$ satisfies
\eqn{framesymB3}{
g(e_0,e_J)= 0 \AND F\bigl( (-g(e_0,e_0))^{-1/2}\bigr)^2 = \det(g(e_I,e_J))
}
on $V_T$, and defines a $C^1$ solution of the boundary value problem
\lalign{framesymB4}{
A_{\nu\mu}{}^\gamma \nabla_\gamma e^\mu_l &= -2 g(e_0,e_0) \theta^i_\nu \delta^0_{(i}\delta^k_{j)}\sigma_l{}^j{}_k,
\label{framesymB4.1}\\
e_0(\sigma_l{}^j{}_k) & = 0, \label{framesymB4.2}\\
g(e_0,e_0)|_{\Gamma_T} & = c_0, \label{framesymB4.3}\\
g(e_0,e_J)|_{\Gamma_T} & = 0, \label{framesymB4.4}
}
on $V_T$ where
\leqn{framesymB5}{
\sigma_l{}^j{}_k = \theta^j([e_l,e_k]) = \theta^j_\lambda
\bigl( e^\sigma_l \del{\sigma} e^\lambda_k  - e^\sigma_k \del{\sigma} e^\lambda_l \bigr).
}
\end{prop}

To prove the converse of this Proposition, we adapt the proof of Proposition \eqref{equivA}. However,
unlike the previous Proposition, the proof of \eqref{equivA} does not immediately adapt to the boundary
setting because in the proof of Proposition \eqref{equivA} we used the uniqueness property of
solutions to symmetric hyperbolic systems, and this property depends strongly on the boundary conditions when a boundary
is present.

\begin{prop} \label{equivB}
Suppose $T>0$, $g_{\mu\nu}\in C^2(V_T)$, and
$\{\{e_i\}_{i=0}^3,\sigma_l{}^j{}_k\}$ is a $C^2$ solution of the boundary value problem \eqref{framesymB4.1}-\eqref{framesymB4.4} that
satisfies the initial data constraints
\lgath{equivB1}{
[e_0,e_I]|_{\Sigma_0} = 0, \quad g(e_0,e_I)|_{\Sigma_0} = 0, \quad
\det\bigl(g(e_I,e_J)|_{\Sigma_0}\bigr)= F\bigl((-g(e_0,e_0)|_{\Sigma_0})^{-\frac{1}{2}}\bigr) \label{equivB1.1}
\intertext{and}
\sigma_l{}^j{}_k|_{\Sigma_0} = \bigl( \theta^j([e_l,e_k]) \bigr)|_{\Sigma_0} = \Bigl(\theta^j_\lambda
\bigl( e^\sigma_l \del{\sigma} e^\lambda_k  - e^\sigma_k \del{\sigma} e^\lambda_l \bigr)\Bigl)\bigl|_{\Sigma_0}.
\label{equivB1.2}
}
Then $\{e_i\}_{i=0}^3$ with $w=e_0$ defines a $C^2$ solution of \eqref{framesymB1.1}-\eqref{framesymB1.3} on $V_T$.
\end{prop}
\begin{proof}
Let $w=e_0$, and suppose that $\{e_i\}_{i=0}^3$ is a $C^2$ solution of \eqref{framesymB4.1}-\eqref{framesymB5} and that
the constraints on the initial data \eqref{equivA1.1} are satisfied. Next, we observe that
\leqn{equivB2}{
\sigma_0{}^j{}_k = \sigma_k{}^j_0 = 0
}
follows from the evolution equation \eqref{framesymB4.2} and the
 fact that the choice of initial data \eqref{equivB1.1}-\eqref{equivB1.2} implies that
 $\sigma_0{}^j{}_k|_{\Sigma_0} = \sigma_k{}^j{}_0|_{\Sigma_0} = 0$.
Substituting \eqref{equivB2} into the evolution equation \eqref{framesymB4.1} yields
\lalign{equivB3}{
A_{\nu\mu}{}^\gamma \nabla_\gamma e_0^\mu &= 0, \label{equivB3.1}\\
g(e_0,e_0)|_{\Gamma_T} & = c_0, \label{equivB3.2}
}
or in other words, the vector field $w = e_0$ satisfies the EFW equations \eqref{eul1} and
the boundary condition $p|_{\Gamma_T} = 0$.
Setting $\ech_0= e_0$, we can propagate the initial data $e_I|_{\Sigma}$ by the flow of $\ech_0$ to
get a $C^1$ solution $\{\ech_I\}_{I=1}^3$ on $V_T$ of the initial boundary value problem
\lalign{equivB4}{
\Achk_{\nu\mu}{}^\gamma \nabla_\gamma \ech_0 & = 0, \label{equivB4.1} \\
[\ech_0,\ech_I] & = 0, \label{equivB4.2}\\
g(\ech_0,\ech_0)|_{\Gamma_T} & = c_0 \label{equivB4.3}\\
\ech_I|_{\Sigma_0} & = e_I|_{\Sigma_0}\label{equivB4.4}
}
where
\leqn{equivB5}{
\Achk_{\nu\mu}{}^\gamma = \left(3 + \frac{1}{\sch^2}\right) \frac{\wch_\mu \wch_\nu}{\wch^2} \wch^\gamma + \delta^\gamma_\nu \wch_\mu
+\delta^\gamma_\mu \wch_\nu + \wch^\gamma g_{\mu\nu},
}
$\wch = \ech_0$, $\sch^2 = s^2((\wch^2)^{-1/2})$  and $\wch^2 = \wch_\mu \wch^\mu$. By definition $\wch = \ech_0 = e_0 =w$, and thus, we have that
\leqn{equivB6}{
\Achk_{\nu\mu}{}^\gamma = A_{\nu\mu}{}^\gamma
}
where $A_{\nu\mu}{}^\gamma$ is as previously defined (see \eqref{eul3}).

Since $\ech_0 = w$, it follows from \eqref{equivB1.1}, \eqref{equivB4.1}-\eqref{equivB4.4} and Proposition \ref{framesymB}
that $\{\ech_i\}_{i=0}^3$ defines a $C^1$ solution of the initial boundary value problem
\lalign{equivB7}{
\Achk_{\nu\mu}{}^\gamma \nabla_\gamma \ech^\mu_l &= -2 g(\ech_0,\ech_0) \thetach^i_\nu \delta^0_{(i}\delta^k_{j)}\sigmach_l{}^j{}_k,
\label{equivB7.1}\\
\ech_0(\sigmach_l{}^j{}_k) & = 0, \label{equivB7.2}\\
g(\ech_0,\ech_0)|_{\Gamma_T} & = c_0, \label{equivB7.3}\\
g(\ech_0,\ech_J)|_{\Gamma_T} & = 0, \label{equivB7.4}\\
\ech_j|_{\Sigma_0} & = e_j|_{\Sigma_0}, \label{equivB7.5}\\
\sigmach_l{}^j{}_k|_{\Sigma_0} & = \bigl( \theta^j([e_l,e_k]) \bigr)|_{\Sigma_0} \label{equivB7.6}
}
where $\{\thetach^i\}_{j=0}^3$ is the basis dual to $\{\ech_j\}_{j=0}^3$
and $\sigmach_l{}^j{}_k = \thetach^j([\ech_l,\ech_k])$.

From \eqref{framesymB4.2}, \eqref{equivB1.2}, \eqref{equivB7.2}, \eqref{equivB7.6}, and $e_0=\ech_0$, we see that
\eqn{sigmadiffA}{
e_0(\sigma_l{}^j{}_k-\sigmach_l{}^j{}_k) = 0 \AND \bigl(\sigma_l{}^j{}_k-\sigmach_l{}^j{}_k\bigr)|_{\Sigma_0} = 0,
}
and so we conclude that
\leqn{sigmadiffB}{
\sigma_l{}^j{}_k = \sigmach_l{}^j{}_k.
}
Setting
\eqn{fdef}{
f^\mu_j = e^\mu_j - \ech^\mu_j,
}
it follows from \eqref{equivB4.1}-\eqref{equivB4.4}, \eqref{equivB6}, \eqref{equivB7.1}, \eqref{equivB7.3}-\eqref{equivB7.4},
\eqref{sigmadiffB}
and $\ech_0= e_0$ that the $f^\mu_j$ define a $C^1$ solution of the initial boundary value problem
\lalign{equivB8}{
A_{\nu\mu}{}^\gamma \nabla_\gamma f^\mu_l &= B_{\nu\mu}{}_l^{m} f^\mu_m,
\label{equivB8.1}\\
g(e_0,f_j)|_{\Gamma_T} & = 0, \label{equivB8.2}\\
f_j|_{\Sigma_0} & = 0 \label{equivB8.3}
}
on $V_T$ where the coefficients $B_{\nu\mu}{}_l^m$ are $C^1$ on $V_T$.

We can use the time parameter $t$ of the flow $\Fc_t$ of $e_0$ to introduce a
time function $t$ that satisfies
$\Ip_{e_0} \ed t = 1$,
and whose level sets are $\Sigma_t$ for $t\in (0,T)$. We also observe that the boundary of $V_t$ decomposes
as
\eqn{equivB10}{
\partial V_t = \Sigma_t \cup \Sigma_0 \cup \Gamma_t.
}
Letting
\eqn{equivB11}{
t_\mu = \frac{1}{\sqrt{-g^{\alpha\beta}\del{\alpha}t\del{\beta}t}}\del{\mu} t,
}
we see that $-t_\mu|_{\Sigma_t}$
is the inward pointing unit co-normal to $\Sigma_t$, and
$t_\mu|_{\Sigma_0}$
is the inward pointing unit co-normal to $\Sigma_0$. We let $n_\mu$ be a normalized 1-form
such that $n_\mu|_{\Gamma_t}$ is a unit outward pointing co-normal to $\Gamma_t$.
Since $w=e_0|_{\Gamma_t} \in T \Gamma_t$, we note that
\leqn{equivB13}{
n_\mu w^\mu |_{\Gamma_t} = 0,
}
which will be used in an essential manner below.

As discussed in Remark \ref{fraurem}, the bilinear form
$\ipe{f}{k} = - t_\gamma A_{\mu\nu}{}^\gamma \delta^{ij} f^{\mu}_i k^{\nu}_j$
defines positive definite inner-product. This together with the index symmetry $A_{\mu\nu}{}^\gamma = A_{\nu\mu}^\gamma$ implies
that the equation \eqref{equivB8.1} is symmetric hyperbolic, and consequently we have
an energy estimate of the form
\leqn{equivB15}{
\int_{\Sigma_t} \ipe{f}{f}\, d \mu_t + \int_{\Gamma_t} n_\gamma A_{\mu\nu}{}^\gamma \delta^{ij} f^{\mu}_i f^{\nu}_j \,d\nu_t
= \int_{\Sigma_0} \ipe{f}{f}\, d \mu_0 + \int_{V_t} \delta^{ij}\bigl(f^\nu_i B_{\nu\mu}{}_j^m f_m^\mu
+ f^\mu_i\nabla_\gamma A_{\mu\nu}^\gamma f^{nu}_j\bigr)\, dV
}
where $dV$ is the volume element determined by the metric $g$, and $d\mu_t$ and $d\nu_t$ are the
induced volume elements on $\Sigma_t$ and $\Gamma_t$, respectively.

Letting
\eqn{equivB16}{
h_{\mu\nu} = g_{\mu\nu} + \frac{w_\mu w_\nu}{w^2}
}
denote the positive definite metric on the subspace orthogonal to $w^\mu$, we can write $A_{\mu\nu}{}^\gamma$
as (see \eqref{eul3})
\eqn{equivB17}{
A_{\mu\nu}^\gamma = \frac{1}{s^2 w^2}w_\mu w_\nu w^\gamma + h^\gamma{}_\nu w_\mu + h^\gamma{}_\mu w_\nu
+ w^\gamma h_{\mu\nu}.
}
Using this, \eqref{equivB13}, and $w=e_0$, we then get that
\eqn{equivB19}{
n_\gamma A_{\mu\nu}{}^\gamma \delta^{ij} f^{\mu}_i f^{\nu}_j |_{\Gamma_t} = \bigl(2\delta^{ij} f^\nu_i n_\gamma h^\gamma{}_\nu
g(e_0,f_j)\bigr)|_{\Gamma_t}.
}
Applying the boundary condition \eqref{equivB8.2}, we arrive at the conclusion
\eqn{equivB20}{
n_\gamma A_{\mu\nu}{}^\gamma \delta^{ij} f^{\mu}_i f^{\nu}_j |_{\Gamma_t} = 0.
}
Substituting this into the energy estimate \eqref{equivB15} then yields
\eqn{equivB21}{
\int_{\Sigma_t} \ipe{f}{f} \, d \mu_t
= \int_{\Sigma_0} \ipe{f}{f}\, d \mu_0 + \int_{V_t} \delta^{ij}\bigl(f^\nu_i B_{\nu\mu}{}_j^m f_m^\mu
+ f^\mu_i\nabla_\gamma A_{\mu\nu}^\gamma f^{nu}_j\bigr)\, dV.
}
From this, the Cauchy-Shwartz inequality and the fact that \eqref{equivB14} defines a positive definite inner-product,
it follows that there exists a constant $C>0$ such that
\eqn{equivB22}{
\int_{\Sigma_t} \ipe{f}{f}\, d \mu_t
\leq \int_{\Sigma_0} \ipe{f}{f}\, d \mu_0 +  C\int_0^t \int_{\Sigma_s} \ipe{f}{f}\, d\mu_s\, dt
}
for all $t\in [0,T]$. Applying Growall's inequality, we find
\eqn{equivB23}{
\int_{\Sigma_t} \ipe{f}{f}\, d\mu_t \leq e^{Ct} \int_{\Sigma_0} \ipe{f}{f} \,d \mu_0.
}
A direct consequence of this inequality and the initial condition \eqref{equivB8.3}  is that
$f^\mu_j = 0$ on $V_T$. Therefore, we conclude
that $\ech^\mu_j = e^\mu_j$ and the proof is complete.
\end{proof}

\subsubsect{Jacsym}{A symmetric hyperbolic equation for the inverse Jacobian matrix}

Returning to the problem of deriving an evolution equation for the inverse Jacobian matrix, we
claim that the equation \eqref{fwev21.1} for the frame fields $e^\mu_i$ when evaluated in Lagrange coordinates
can be used to control the inverse Jacobian matrix \eqref{Jinv}.
To see this, we first make the following definitions:
\lalign{tvarsdef}{
\et^\mu_j &:= e^\mu_j \circ \phi, \label{tvarsdef.1}\\
\thetat^j_\mu &:= \theta^j_\mu\circ \phi, \label{tvarsdef.2}\\
\gt_{\mu\nu} &:= g_{\mu\nu}\circ\phi, \label{tvarsdef.3}\\
\gt^{\mu\nu} &:= g^{\mu\nu}\circ \phi, \label{tvarsdef.4}\\
\gt_{\gamma\nu\nu} &:= g_{\gamma\mu\nu}\circ \phi, \label{tvarsdef.5}\\
\Gammat^\gamma_{\mu\nu} &:= \Gamma^\gamma_{\mu\nu}\circ \phi \notag \\
&=  \Half \gt^{\gamma\sigma}\bigl(\gt_{\mu\nu\sigma} + \gt_{\nu\mu\sigma}- \gt_{\sigma\mu\nu}\bigr), \label{tvarsdef.6}\\
\sigmat_l{}^k{}_j &:= \sigma_l{}^k{}_j\circ \phi, \label{tvarsdef.7}
\intertext{and}
\At_{\mu\nu}{}^\gamma &:= A_{\mu\nu}{}^\gamma \circ \phi \notag \\
&=  \left(3 + \frac{1}{\sbar^2}\right)
\frac{\gt_{\mu\lambda}\gt_{\nu\sigma} \et_0^\lambda \et_0^\sigma}{\wb^2} \et^\gamma_0 + \delta^\gamma_\nu \gt_{\mu\lambda} \et^\lambda_0
+\delta^\gamma_\mu \gt_{\nu\lambda}\et^\lambda_0 + \et_0^\gamma \gt_{\mu\nu} \label{tvarsdef.8}
\intertext{where}
\sbar^2 &:= s^2\left(\frac{1}{\wb}\right) \quad \text{with} \quad \wb := \sqrt{-\gt_{\mu\nu}\et^\mu_0 \et^\nu_0}. \label{tvarsdef.9}
}
We also note that
\leqn{tvars1}{
(\gt^{\mu\nu}) = (\gt_{\mu\nu})^{-1} \AND (\thetat^j_\mu) = (\et^\mu_j)^{-1}.
}

\begin{rem} \label{lagrem}
It is worthwhile remarking that in our Lagrange formulation we take as our primary
variables the components of the the geometric objects with respect to harmonic coordinates, i.e. $\{e^j_\mu$, $g_{\mu\nu}$, $\ldots\}$,
evaluated in the Lagrange coordinates determined by the map $\phi$, i.e. $\{\et^j_\mu = e^j_\mu\circ \phi$, $\gt_{\mu\nu} = g_{\mu\nu}\circ \phi$, $\ldots\}$.
This construction can be geometrized using the
(functorial) pull back bundle construction, which coincides in local coordinates with the operation of leaving the components
of a geometric object untransformed while at the same time changing the point where the components are evaluated by composition
with a map. However, we will not do this here as it does not add anything essential to the arguments below.

We also observe that given the variables $\{\et^j_\mu$, $\gt_{\mu\nu}$, $\ldots\}$,
the original ones $\{e^j_\mu$, $g_{\mu\nu}$, $\ldots\}$ can be recovered by
composition with $\phi^{-1}$.  As we show below, $\phi$, and consequently, $\phi^{-1}$, is determined
uniquely in terms of the  $\{\et^j_\mu$, $\gt_{\mu\nu}$, $\ldots\}$ variables
by a differential equation, see \eqref{diffev} below.
\end{rem}

Using the definitions \eqref{tvarsdef.1}-\eqref{tvarsdef.9}, we can write the evolution equation
\eqref{fwev21.1} as
\leqn{fwev23}{
\At_{\nu\mu}{}^\gamma \Jch^\lambda_\gamma \delb{\lambda} \et^\mu_l =
- \At_{\nu\mu}{}^\gamma \Gammat_{\gamma \lambda}^\mu \et^\lambda_l + 2 \wb^2 \thetat^i_\nu \delta^0_{(i}\delta^k_{j)}\sigmat_l{}^j{}_k
}
where $\sigmat_l{}^k{}_j$ satisfies\footnote{We note that (see \eqref{fwev21.2} ) $\sigma_l{}^k{}_j$ satisfies
$e_0(\sigma_l{}^k{}_j)=0$. From this and the change of variable formula, it follows that $\sigmat_l{}^k{}_j = \sigma_l{}^k{}_j\circ \phi$
satisfies \eqref{fwev24}.}
\leqn{fwev24}{
\delb{0}\sigmat_l{}^k{}_j = 0.
}

Next, we consider the pullback of the frame $e_i$ by $\phi$:
\leqn{fwev25}{
\eb_i = \phi^*\eb_i \quad \Longleftrightarrow  \quad \eb_i^\mu = \Jch^\mu_\nu \et^\nu_j,
}
which satisfy
\leqn{fwev27}{
[\eb_0,\eb_j] = 0 \quad \Longleftrightarrow \quad \delb{0} \eb^\mu_j = 0
}
by\footnote{We recall that the Lie bracket $[\cdot,\cdot]$ satisfies $\psi^*[X,Y]=[\psi^*X,\psi^*Y]$ for
all diffeomorphisms $\psi$ and vector fields $X,Y$.} \eqref{eIdef}.

By definition (see \eqref{Sigma} and \eqref{phidefA}), the
diffeomorphism $\phi$ satisfies
\leqn{phiSigma}{
\phi^{-1}(\Sigma) = \Sigma,
}
and this, in turn, implies that
\leqn{etidata}{
\et^\mu_j|_{\Sigma} = e^\mu_j|_{\Sigma}.
}
It also follows from \eqref{wbdefB}, \eqref{e0def}, \eqref{tvarsdef.1}, \eqref{phiSigma}, and \eqref{etidata} that
\leqn{Jidata}{
J^\mu_I|_{\Sigma} = \delta^\mu_I \AND J^\mu_0|_{\Sigma} = e^\mu_0|_{\Sigma}.
}
Now, \eqref{tvars1} and \eqref{fwev25}-\eqref{etidata} imply that
\leqn{Jchev}{
\Jch^\mu_\nu = \Jc^\mu_i \thetat^i_\nu
}
where
\leqn{Jcdef}{
\Jc^\mu_i = \Jch^\mu_\lambda|_{\Sigma} e^\lambda_i|_{\Sigma}.
}
We note that
\leqn{Jchidata}{
\bigl(\Jch^\mu_\nu|_\Sigma\bigr) = \begin{pmatrix}\begin{displaystyle} \frac{1}{e^0_0|_{\Sigma}}\end{displaystyle} & 0\\
\begin{displaystyle} -\frac{1}{e^0_0|_{\Sigma}}e^\Lambda_0|_{\Sigma}\end{displaystyle} & \id_{3\times 3}\end{pmatrix}
}
by \eqref{Jidata}. Together, equations \eqref{fwev23}, \eqref{fwev24} and \eqref{Jchev} form a symmetric hyperbolic
system that completely determines the inverse Jacobian matrix thereby justifying our claim that equation \eqref{fwev21.1} when transformed into the Lagrange coordinates
can be used to control the inverse Jacobian matrix.

\subsect{lagEE}{The Einstein-Euler equations in Lagrange coordinates}

Using the definitions \eqref{tvarsdef.1}-\eqref{tvarsdef.9} and \eqref{tvars1}-\eqref{fwev24}, and
\eqref{Jchev}- \eqref{Jchidata} of the
previous section, the Einstein-Euler equations \eqref{EEc.1}-\eqref{EEc.3} in the Lagrangian coordinates become
\lalign{EElag}{
\Bt^{\alpha \beta \gamma}\Jc^\lambda_m \thetat^m_\gamma \delb{\lambda} \gt_{\beta\mu\nu} &= -\et_0^\alpha\bigl(Q_{\mu\nu}(\gt_{\tau\delta},
\gt_{\sigma\tau\delta})-2\kappa\bigl(\Tt_{\mu\nu}-\Half \Tt \gt_{\mu\nu}\bigr)\bigr) \label{EElag.1}, \\
\At_{\nu\mu}{}^\gamma \Jc^\lambda_m \thetat^m_\gamma \delb{\lambda} \et^\mu_l &=
- \At_{\nu\mu}{}^\gamma \Gammat_{\gamma \lambda}^\mu \et^\lambda_l + 2 \wb^2 \thetat^i_\nu \delta^0_{(i}\delta^k_{j)}\sigmat_l{}^j{}_k, \label{EElag.2}\\
 \delb{0} \gt_{\mu\nu} &= \et_0^\gamma \gt_{\gamma\mu\nu}, \label{EElag.3}
\intertext{and}
\delb{0} \sigmat_l{}^j{}_k &= 0 \label{EElag.4}
}
where
\alin{BTtdef}{
\Bt^{\alpha \beta \gamma} &= -\et_0^\alpha \gt^{\beta \gamma} - \et_0^\beta \gt^{\alpha\gamma}
+ \et_0^\gamma \gt^{\alpha\beta},\\
\Tt^{\mu\nu} &= \left(\rho\left(\frac{1}{\wb}\right)+p\left(\frac{1}{\wb}\right)\right)
\frac{\et_0^\mu \et_0^\nu}{\wb^2} + p\left(\frac{1}{\wb}\right)\gt^{\mu\nu},
\intertext{and}
\Tt &= \gt_{\mu\nu}\Tt^{\mu\nu}.
}
We also note that the diffeomorphism $\phi=(\phi^\mu)$ can be determined by integrating the equations
\leqn{diffev}{
\delb{0}\phi^\mu = \et^\mu_0.
}

The system \eqref{EElag.1}-\eqref{EElag.4} is symmetric hyperbolic and initial data for this system
is given by
\eqn{EElagidata}{
\bigl(\gt_{\mu\nu}|_{\Sigma},\gt_{\gamma\mu\nu}|_\Sigma,\et^\mu_j|_\Sigma,\sigmat_l{}^k{}_j|_{\Sigma}\bigr)=
\left(g_{\mu\nu}|_{\Sigma},\del{\gamma}g_{\mu\nu}|_\Sigma,e^\mu_j|_\Sigma, \Bigl(\theta^j_\lambda
\bigl( e^\sigma_l \del{\sigma} e^\lambda_k  - e^\sigma_k \del{\sigma} e^\lambda_l \bigr)\Bigl)\bigl|_{\Sigma}\right)
}
where
\begin{itemize}
\item[(i)] $w^\mu|_\Sigma = e_0^\mu|_\Sigma $ satisfies the restrictions \eqref{widata} and \eqref{irho},
\item[(ii)] $\bigl(g_{\mu\nu}|_\Sigma,\del{\gamma}g_{\mu\nu}|_{\Sigma}\bigr)$ satisfies the constraint
equations \eqref{ceqns.1}-\eqref{ceqns.2}, and
\item[(iii)] the $e^\mu_I|_\Sigma$ are chosen so that $e^\mu_j|_{\Sigma}$ defines a linearly independent set of vectors,
and the restrictions \eqref{eIidata} and $[e_0,e_I]|_{\Sigma}=0$ are satisfied.
\end{itemize}

\subsubsect{eelag}{Equivalence of \eqref{EEc.1}-\eqref{EEc.3} and \eqref{EElag.1}-\eqref{EElag.4}}

Thus far, the arguments of the above sections show that sufficiently smooth solutions of the $1^{\text{st}}$ order form of the harmonically
reduced Einstein-Euler equations \eqref{EEc.1}-\eqref{EEc.3} satisfy \eqref{EElag.1}-\eqref{EElag.4}
when evaluated in Lagrange coordinates. We now claim that the converse of this statement
follows immediately from the uniqueness of solutions to symmetric hyperbolic systems. To see this, start with initial data
\eqn{eelag1}{
Z|_\Sigma = \bigl(g_{\gamma \mu \nu}|_{\Sigma} = \del{\gamma}g_{\mu\nu}|_{\Sigma}, g_{\mu\nu}|_{\Sigma}, e^\mu_j|_{\Sigma}\bigr)
}
satisfying restriction (i)-(iii) above. Assume that $Z|_{\Sigma}$ is chosen sufficiently smooth so that when it is evolved
using \eqref{EEc.1}-\eqref{EEc.3}, \eqref{e0def} and \eqref{eIdefa}, the resulting solution
\eqn{eelag2}{
Z = \bigl(g_{\gamma \mu \nu}, g_{\mu\nu}, e^\mu_j\bigr)
}
is sufficiently smooth, say $C^2$. We also note that it is well known that solutions of the
$1^{\text{st}}$ order formulation \eqref{EEc.1} of the Einstein equations with initial data $ g_{\gamma \mu \nu}|_{\Sigma} = \del{\gamma}g_{\mu\nu}|_{\Sigma}$
satisfy
\eqn{eelag2a}{
g_{\gamma \mu \nu} = \del{\gamma}  g_{\mu \nu}.
}

Integrating $e^\mu_0$ to get the flow map \eqref{flowA} and defining the diffeomorphism
$\phi$ as before (see \eqref{phidefA}), the arguments of the preceding sections show that
\eqn{eelag3}{
\tilde{Z}_1 =  \bigl(g_{\gamma \mu \nu}\circ \phi = (\del{\gamma}g_{\mu\nu})\circ \phi , g_{\mu\nu}\circ \phi, e^\mu_j \circ \phi,  \sigma_l{}^j{}_k \circ \phi \bigr)
}
defines a $C^1$ solution of \eqref{EElag.1}-\eqref{EElag.4} with initial data
\leqn{eelag4}{
\tilde{Z}|_{\Sigma} = \bigl(g_{\gamma \mu \nu}|_{\Sigma}=\del{\gamma}g_{\mu\nu}|_{\Sigma}, g_{\mu\nu}|_{\Sigma}, e^\mu_j|_{\Sigma},\sigma_l{}^j{}_k\bigr|_\Sigma).
}
Next let
\eqn{eelagb}{
\tilde{Z}_{2} = \bigl(\gt_{\gamma \mu \nu}, \gt_{\mu\nu}, \et^\mu_j,\sigmat_l{}^j{}_k\bigr)
}
be the $C^1$ solution of \eqref{EElag.1}-\eqref{EElag.4} generated by the same initial data \eqref{eelag4}. Then by the uniqueness of
$C^1$ solutions of symmetric hyperbolic systems, we must have that
\eqn{eelag5}{
\tilde{Z}_1 = \tilde{Z}_2.
}
Thus solutions of \eqref{EElag.1}-\eqref{EElag.4} with initial data satisfying restrictions (i)-(iii) above determine solutions
of the Einstein-Euler equations, and in particular, the relation
\eqn{eelag5a}{
\gt_{\gamma \mu \nu} = (\del{\gamma}  g_{\mu \nu})\circ \phi
}
holds for such solutions. 
\sect{zs}{Densitized lapse and zero shift coordinate fixing}

As an application, we now consider fictitious Lagrange
coordinates for the vacuum Einstein equations. These coordinates, as we show below, are equivalent
to a densitized lapse and zero shift coordinate system. We note that a similar application of
fictitious Lagrange coordinates has already been pointed out in \cite{Frie}.

We start by again assuming
\leqn{kappazero}{
\kappa = 0
}
so that we are dealing with the vacuum Einstein equations in Lagrange coordinates.
Next, we
specify the metric initial data
\eqn{zsidataA}{
\bigl(g_{\mu\nu}|_{\Sigma},\del{\gamma}g_{\mu\nu}|_{\Sigma}\bigl)
}
by demanding that in addition to satisfying the gravitational and harmonic constraint equations \eqref{ceqns.1}-\eqref{ceqns.2},
it also satisfies the following densitized lapse and zero shift conditions
\leqn{zsidataB}{
\sqrt{-g_{00}|_\Sigma}= \frac{1}{F^{-1}(\det\bigl(g_{\Lambda \Gamma}|_{\Sigma}\bigr)} \AND g_{0\Lambda}|_\Sigma = 0 \qquad \Lambda,\Gamma = 1,2,3.
}
Here, we observe that the inverse of the function $F(\zeta)$ defined by \eqref{FdefA} is well defined due
to the strict monotonicity of $F(\zeta)$ which follows directly from the formula \eqref{FdefA},  and the positivity of
$s^2$
and $\zeta = 1/w$. For the frames $e^\mu_i$, we choose initial data satisfying
\leqn{zsidataC}{
e^\mu_i|_{\Sigma} = \delta^\mu_i \AND [e_{0},e_I]|_{\Sigma} = 0,
}
which means by way of \eqref{zsidataB} that the restriction on the initial data for
the frames discussed at the end of Section \ref{lagEE} are fulfilled.
Next, we observe that \eqref{zsidataC} implies
\eqn{zsidataD}{
\sigmat_l{}^j{}_{k}|_{\Sigma}  = 0,
}
and hence, by \eqref{EElag.4}, that
\leqn{zsevA}{
\sigmat_l{}^j{}_k = 0.
}
We also observe that
\leqn{Jchzs}{
\Jch^\mu_\nu = \delta^\mu_i\thetat^i_\nu \AND \delb{\mu}\phi^\nu = \delta_\mu^i\et^\nu_i
}
by \eqref{Jchev}, \eqref{Jchidata} and \eqref{zsidataC}. Using \eqref{kappazero}, \eqref{zsevA}, and \eqref{Jchzs}, the evolution
equations \eqref{EElag.1}-\eqref{EElag.4} reduce to
\lalign{Elag}{
\Bt^{\alpha \beta \gamma} \delta^\lambda_i\thetat^i_\gamma \delb{\lambda} \gt_{\beta\mu\nu} &= -\et_0^\alpha Q_{\mu\nu}(\gt_{\tau\delta},
\gt_{\sigma\tau\delta}), \label{Elag.1} \\
\At_{\nu\mu}{}^\gamma \delta^\lambda_i\thetat^i_\gamma \delb{\lambda} \et^\mu_l &=
- \At_{\nu\mu}{}^\gamma \Gammat_{\gamma \lambda}^\mu \et^\lambda_l, \label{Elag.2}
\intertext{and}
 \delb{0} \gt_{\mu\nu} &= \et_0^\gamma \gt_{\gamma\mu\nu}. \label{Elag.3}
}

Letting $\gb_{\mu\nu}$ denote the metric in the $(\xb^\mu)$ coordinates,
we have by \eqref{Jchzs} that
\eqn{gbtransA}{
\gb_{\mu\nu} = \delb{\mu}\phi^\delta \gt_{\delta\tau} \delb{\nu}\phi^\tau,
}
and hence that
\eqn{gbtransB}{
\gb_{\mu\nu} =   \delta^i_\mu \et^\delta_i \gt_{\delta\tau} \et^\tau_j \delta^j_\nu = \delta^i_\mu \delta^j_\mu g_{ij}\circ\phi.
}
But this result combined with \eqref{fwev11} shows that
\eqn{gbzs}{
\sqrt{-\gb_{00}}= \frac{1}{F^{-1}\bigl(\det(\gb_{\Lambda \Gamma})\bigr)} \AND \gb_{0\Lambda} = 0.
}
This relation justifies our assertion that our fictitious Lagrange coordinates are equivalent to a zero-shift and densitized lapse
coordinate condition for the metric.

\begin{rem}
The system \eqref{Elag.1}-\eqref{Elag.3} represents a new densitized lapse and zero shift formulation of the vacuum Einstein equations 
that admits a well-posed initial value problem. As discussed in \cite{NOR2004}, it was shown that, on one hand, the standard ADM formulation of the vacuum Einstein
equations with a densitized lapse and zero shift coordinate condition does not admit a well-posed initial value problem as it is only weakly
hyperbolic,
while, on the other hand, a BSSN-type formulation does admit a well-posed initial value problem with the same gauge choice. From this, one can conclude
that well-posedness of the initial value problem for the vacuum Einstein equations
in a densitized lapse and zero shift gauge is sensitive to the formulation used. For other formulations of the vacuum Einstein
equations that admit a well-posed initial value problem in densitized lapse and zero shift coordinates, see 
\cite{SarTig}.
\end{rem}
\sect{disc}{Discussion}

In this article, we have introduced a new symmetric hyperbolic formulation of the
Einstein-Euler equations in Lagrange coordinates that are adapted to the Frauendiener-Walton formulation
of the Euler equations. This new symmetric hyperbolic formulation in Lagrange coordinates has the
advantage over previous symmetric hyperbolic formulations in Lagrange coordinates of not requiring the use
of higher derivatives of the gravitational and fluid variables in order to formulate a symmetric hyperbolic system.
In principle, this should make the boundary behavior of our system easier to analyze.

While we do not have any definitive existence results in the presence of a boundary,
the results of Section \ref{bequiv} do show that our formulation does possess desirable
properties that are likely to be useful in analyzing the free boundary problem.
We are currently extending the results of Section \ref{bequiv} with the aim of
formulating the free boundary value problem as a standard boundary value problem on
a fixed domain where standard techniques can be used to prove existence and uniqueness.
We are optimistic that this can be done in the situation where there is no coupling to
gravity, that is where the metric is considered as a fixed flat metric.
However, when gravity is present, we have not been able to identify a promising formulation of the combined gravitational
fluid system with a natural set of boundary conditions. Clearly, new ideas are needed to address the free
boundary problem for the full Einstein-Euler equations.

When there is no boundary, we believe that our new Lagrange formulation is still interesting
due to the fact that our Lagrange coordinates correspond to densitized lapse and zero shift
coordinate gauge where the functional dependence of the lapse on the determinant of the spatial
three metric is determined by the equation of state of the fluid. As we have shown in Section \ref{zs},
this construction is even interesting for the vacuum Einstein equations as it allows us to construct a
wide class of densitized lapse and zero shift gauges by introducing Lagrange coordinates
adapted to a fictitious fluid. 

\begin{acknowledgments}
This work was partially supported by the ARC grant DP1094582 and a MRA grant. Part of this work was competed while visiting
the Albert-Einstein-Institute (AEI). I thank the AEI and the director Gerhard Huisken of the Geometric Analysis and Gravitation Group for
supporting this research. I also thank the referees for their comments and criticisms which have served to improve the content and exposition
of this article.
\end{acknowledgments}

\bibliography{lagref}	

\end{document}